\documentclass[graybox]{svmult}

\pdfoutput=1

\usepackage{mathptmx}       
\usepackage{helvet}         
\usepackage{courier}        
\usepackage{type1cm}        
\usepackage{makeidx}         
\usepackage{amsmath}         
\usepackage{graphicx}        
\usepackage{multicol}        
\usepackage[bottom]{footmisc}

\makeindex             

\def\wig#1{\mathrel{\hbox{\hbox to 0pt{%
           \lower.5ex\hbox{$\sim$}\hss}\raise.4ex\hbox{$#1$}}}}

\def\Vxc{V_{\rm xc}^{ee}}
\def\mueid{\mu_e^{\rm id}}
\def\nebar{{\bar n}_e^0}
\def\nepa{n_e^{\rm PA}}
\def\neion{n_e^{\rm ion}}
\def\nescr{n_e^{\rm scr}}
\def\neext{n_e^{\rm ext}}
\def\Zbar{{\bar Z}}

\begin{document}

\title*{The structure of warm dense matter modeled with an average atom model with ion-ion correlations}

\titlerunning{Average atom model with ion-ion correlations} 
\author{D. Saumon, C.E. Starrett, J.A. Anta, W. Daughton and G. Chabrier}
\institute{D. Saumon, C.E. Starrett and W. Daughton \at Los Alamos National Laboratory, P.O. Box 1663, Los Alamos, NM 87545, USA  
            \email{dsaumon@lanl.gov, starrett@lanl.gov, daughton@lanl.gov} \and
           J.A. Anta \at Dept. Sistemas Fisicos, Qu\'imicos y Naturales, Universidad Pablo de Olavide, Ctra. Utrera, km 1 - 41013 Sevilla, Spain 
           \email{anta@upo.es}  \and
           G. Chabrier \at CRAL (UMR CNRS 5574), \'Ecole Normale Sup\'erieure de Lyon, 69364 Lyon Cedex 07, France
           \email{chabrier@ens-lyon.fr}}

%
\maketitle

\abstract*{
          We present a new model of warm dense matter that represents an intermediate approach between the relative simplicity of 
          ``one-ion'' average atom models
          and the more realistic but computationally expensive {\it ab initio} simulation methods.  Physical realism is achieved 
          primarily by including the correlations
          in the plasma that surrounds a central ion. The plasma is described with the Ornstein-Zernike integral equations theory of fluids, which is coupled to
          an average atom model for the central ion.  In this contribution we emphasize 
          the key elements and approximations and how they relate to and expand upon a typical
          average atom model. Besides being
          relatively inexpensive computationally, this approach offers several advantages over {\it ab initio} simulations but also has
          a number of limitations.  The model is validated by
          comparisons with numerical solutions for the pair distribution function of the ions from {\it ab initio} simulations for several
          elements and a wide range of plasma conditions.  Simulation results are reproduced remarkably well and 
          simpler limiting theories are recovered as well.  This model has many potential applications to calculation of the properties of
          warm dense matter such as the equation of state and conductivities for a wide range of temperatures and densities.
}
\abstract{
          We present a new model of warm dense matter that represents an intermediate approach between the relative simplicity of 
          ``one-ion'' average atom models
          and the more realistic but computationally expensive {\it ab initio} simulation methods.  Physical realism is achieved 
          primarily by including the correlations
          in the plasma that surrounds a central ion. The plasma is described with the Ornstein-Zernike integral equations theory of fluids, which is coupled to
          an average atom model for the central ion.  In this contribution we emphasize 
          the key elements and approximations and how they relate to and expand upon a typical
          average atom model. Besides being
          relatively inexpensive computationally, this approach offers several advantages over {\it ab initio} simulations but also has
          a number of limitations.  The model is validated by
          comparisons with numerical solutions for the pair distribution function of the ions from {\it ab initio} simulations for several
          elements and a wide range of plasma conditions.  Simulations results are reproduced remarkably well and 
          simpler limiting theories are recovered as well.  This model has many potential applications to calculation of the properties of
          warm dense matter such as the equation of state and conductivities for a wide range of temperatures and densities.
}

\section{Introduction}
\label{sec:intro}
Warm dense matter (WDM) is an ``inner'' frontier of research in material properties in the sense that it is
circumscribed by regimes that can be described by well-understood theoretical approaches and well-known limits. For instance, neither
condensed matter theory, which is primarily based on treating electrons at near-zero temperature in periodic structures, nor plasma
physics theory, based on small Coulomb coupling between the constitutive particles, are applicable to modeling WDM.   
More specifically, WDM is a partially ionized, globally neutral plasma characterized by moderate to strong
coupling between all charged particles (ion-ion, ion-electron, and electron-electron) and partial electron degeneracy,
where weakly bound states can cross into the continuum (pressure ionization). 
Depending on the material, this corresponds to temperatures of $\sim 1-100\,$eV and densities of $\sim 0.1 -50$ times 
the normal solid density.  A general but admittedly vague definition of WDM, is that it occurs in the physical regime where 
``all the physics'' is important. By definition, it is thus quite challenging to model.  

Experimentally, WDM states are not particularly difficult
to achieve nowadays as many experiments cross the WDM regime on their way to denser and, more often, hotter
final states. The difficulty is in making a sample of WDM that is large enough, that is relatively long-lived and with a spatial homogeneity that
allow meaningful probing of its properties. The nature of WDM also limits the effectiveness of well-established diagnostics
of dense matter and plasmas and new methods need to be developed \cite{HEDLP_report}. There is presently only a
limited but growing amount of data on WDM states \cite{desilva98,mancic,eggert08,saiz,kritcher,ciricosta,benuzzi11,dorchies11,ma13}.  

Despite the experimental and theoretical challenges it presents, WDM has long been of interest as it occurs naturally in the giant
planets of the solar system (Jupiter, Saturn, Uranus and Neptune) \cite{guillot05,redmer11}, in dense stars such as in the envelopes of
white dwarf stars (the final stage of a star's life) \cite{fontaine76} and in the many giant and exotic planets 
that have been discovered orbiting other stars in the last decade \cite{fortney09}. In the laboratory, WDM is a transient state of imploding capsules
in inertial confinement fusion experiments and can be produced with pulsed-power platforms and high-power laser facilities. 
Modeling these physical systems typically requires a knowledge of the
equation of state, opacity, conductivity, and diffusion coefficients of WDM.

The intrinsic interest of WDM as a hard physics problem,  as well as its importance to other fields of research has
led to many approximate models that have steadily increased in sophistication.  One large and important class of models,
known as average atoms (AA) models, has a long history and have proved to be very useful. An AA model is
essentially a one-ion model that solves for the electronic structure (bound and free states) around
a central nucleus embedded in a spherically averaged, homogeneous plasma outside the ion sphere. There are many
variants of such models but they all assume spherical symmetry and that the resulting modeled ion, which
typically has a fractional net charge, represents in 
some sense an average of the multiple ion configurations (ionic states) present in the plasma. The assumptions
about the surrounding plasma effectively replace the environment of the central ion by a boundary condition at
the radius of the ion sphere $R$ defined by
\begin{equation}
  \frac{4\pi R^3}{3}=\frac{1}{n_I^0}
  \label{Ris}
\end{equation}
where $n_I^0$ is the number density of ions. The system of equations defining an AA model
(see \S \ref{AA} below) amounts to a self-consistent field problem in spherical symmetry and finite temperature.
This is a conceptually straightforward problem which explains their early and long-lasting popularity 
to model WDM.\footnote{Achieving accurate numerical solutions is more challenging \cite{liberman2}
but has progressed considerably \cite{wilson}.}
Through the boundary condition applied at the ion sphere radius, AA models
can qualitatively, and to a fair extent, quantitatively, predict the electronic structure and charge
of ions as a function of density and temperature \cite{sterne,piron3}.  Despite their popularity and usefulness, average atom
models account for the surrounding plasma only in the simplest possible way and ignore the
correlations in the plasma, an important characteristic of WDM.

At the other end of the spectrum of complexity are computer simulations of WDM, based on {\it ab initio} methods such
as Path Integral Monte Carlo (PIMC) and Quantum Molecular Dynamics (QMD) 
\cite{mazevet1, desjarlais2, zerah, lambert3, ceperley2, driver, holst}.  These methods take a more direct
approach to solving the quantum many-body problem of WDM and use few approximations in simulating the 
properties of a mixture of nuclei and electrons in a simulation box. Bound and free states, radial
and angular correlations,
the formation of clusters, and multiple ion configurations naturally occur in such simulations. The 
theoretical appeal of these methods is tempered by their considerable computational cost
even in view of the computing power available today. Simulations of higher-$Z$ elements remain very onerous.

We present a new model of WDM that takes an intermediate path between simple average atom models
and expensive
simulations, with the goal of producing realistic material properties at a reasonable computational
cost.  The concept of the average atom is extended by including 
correlations between charged particles in the surrounding plasma. The structure of the central ion is thus solved in the field of
the central nucleus and of the surrounding, correlated ions and electrons.  This re-introduces the
surrounding plasma correlations in a much more realistic fashion.
By maintaining spherical symmetry in the formulation of the model, the computational cost 
remains modest.  This opens the possibility of generating large tables of properties, particularly of 
the equation of state, that can be used in various applications.  Similar models have been published in the past 
\cite{xu, chihara91, perrot2, ofer}
but this new model is based on a more formal derivation and a higher level of 
internal consistency.  Limitations of earlier models have been overcome, resulting in a model that is
applicable to a broad range of physical regimes, from liquid metals to WDM and high-temperature plasmas. 

We first briefly review a typical average atom model in \S \ref{AA} to set the context for its extension
to include plasma correlations. The full model is described in \S \ref{aa_tcp}, with illustrations of the
key models quantities. The models has been applied to a wide range of elements, temperature and densities,
and it has been validated by comparisons with {\it ab initio} simulations, primarily in terms of pair distribution
functions (\S \ref{results}). The last section (\S 5) offers a summary and outlook.
This contribution emphasizes the concepts that underlie the model and how it differs from previous similar 
efforts.  A detailed presentation along with derivations are given in \cite{starrett2}, where additional numerical 
results can also be found.  

\section{Average atom models}
\label{AA}
\begin{figure}
  \includegraphics[scale=.7, trim= 20mm 155mm 0mm 15mm, clip]{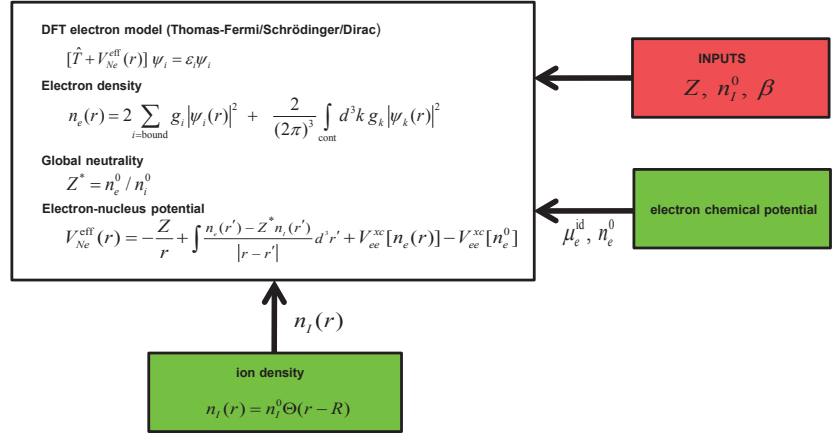}
  \caption{Structure and key equations of a typical average atom model. The red box gives the input parameters to the model. 
           Green boxes show model components where other choices are possible. Arrows show the flow of information between the
           key elements of the model. See text for details.}
  \label{fig:aa_diag}
\end{figure}

An average atom model describes a single central nucleus of charge $Z$ with $Z$ electrons embedded in a
radially and spherically smoothed plasma residing outside the ion sphere surrounding the
nucleus. The surrounding plasma has number density $n_I^0$ and, like the central ion, 
is at a temperature $T=1/\beta$.
Average atom models invariably describe the electron cloud surrounding the nucleus with the finite
temperature formulation of the Kohn-Sham density functional theory (DFT) \cite{kohn, mermin,parr-wang}. The latter
can be cast in the semi-classical Thomas-Fermi model of the electrons \cite{thomas,fermi,feynman}, a full quantum mechanical 
description with the Schr\"odinger equation \cite{blenski2}, or the relativistic quantum mechanical Dirac equation \cite{liberman, wilson, piron3}.
For the clarity of the discussion and without loss of generality, hereafter we will consider only 
the quantum mechanical case (Schr\"odinger equation). The structure of a typical AA model and its key equations
are shown in Fig. \ref{fig:aa_diag}, which is a guide to the following discussion of the model.
The effective one-electron DFT Schr\"odinger equation is
\begin{equation}
  [{\hat T} + V_{Ne}^{\rm eff}(r)] \psi_i(r) = \epsilon_i \psi_i(r)
\end{equation}
where $\hat T$ is the kinetic energy operator, $V_{Ne}^{\rm eff}(r)$ is the effective one-electron potential
energy, $\psi_i(r)$ is the one-electron radial wave function of state $i$ with energy $\epsilon_i$. The outer boundary condition is 
applied at $r \rightarrow \infty$ where $V_{Ne}^{\rm eff} \rightarrow 0$ and a free particle solution is obtained.
$\psi_i(r)\rightarrow 0$ at infinity  for bound states. For continuum states, $\psi_i(r\rightarrow \infty)$ is required to match the  
free solution (spherical Bessel function). In practice, the outer boundary condition is applied at some large radius $R_{\rm max}$
that defines the edge of the computational boundary, where $\psi_i(R_{\rm max}$)=0 for bound states, and $\psi_i(R_{\rm max})$  is
matched to spherical Bessel functions for continuum states \cite{blenski}. The potential is
\begin{equation}
  V_{Ne}^{\rm eff}(r)= -\frac{Z}{r} + \int \frac{n_e(r^\prime)-Z^*n_I(r^\prime)}{|\vec r-\vec r^\prime|}\,d^3r^\prime + \Vxc[n_e(r)] - \Vxc[n_e^0]
  \label{VNe}
\end{equation}
where $n_e(r)$ and $n_I(r)$ are the radial density profiles of electrons and ions around the central nucleus, respectively, 
$Z^*$ is the ion charge in the average atom model,
and $\Vxc$ is the electron exchange and correlation potential.
The electron density is obtained by summing the density associated with the eigenfunctions of both bound and continuum states
\begin{equation}
  n_e(r)= 2\sum_{i={\rm bound}} g_i|\psi_i(r)|^2 + \frac{2}{(2\pi)^3} \int\limits_{\rm cont} d^3k\,g_k|\psi_k(r)|^2
  \label{e_density}
\end{equation}
and the free electron density is
\begin{equation}
  n_e^0=\lim_{r \to \infty} \, n_e(r).
\end{equation}
The $g_i$ and $g_k$ are Fermi occupation factors
\begin{equation}
  g_i= \frac{1}{e^{\beta(\epsilon_i-\mueid)}+1},
  \label{fermi}
\end{equation}
where $\mueid$ is the ideal chemical potential of the electrons. Global neutrality of the plasma requires
that the ion charge be 
\begin{equation}
  Z^*=n_e^0/n_I^0.  
  \label{zstar}
\end{equation}
This essentially defines the self-consistent field problem of the AA model.

Two additional elements are needed to close this system of equations, neither of which is unique (green boxes in 
Fig. \ref{fig:aa_diag}).   One is the density profile of the ions surrounding
the central nucleus, $n_I(r)$, which describes the surrounding plasma.  In AA models, this is taken as a simple step
function with a constant ion density outside the ion sphere:
\begin{equation}
  n_I(r)=n_I^0\Theta(r-R),
  \label{ni}
\end{equation}
which describes a cavity (the ion sphere) in the plasma centered at the origin. This
form was originally devised to approximate the periodic nature of solids in AA models \cite{wigner,xu}, by confining
each average atom within the ion sphere\footnote{In this case the boundary condition is applied at $r=R$.}
 but it can also be interpreted as the ion pair distribution function in a dense plasma.
With this choice for $n_I(r)$, the average atom is coupled to the external plasma only through the ion sphere radius $R$,
which depends only on the density $n_I^0$.  This choice of $n_I(r)$ implies that the external ions are not
correlated with each other or with the central nucleus. It has the virtue of extreme simplicity but is clearly quite approximate
and leaves out important physics. 

The second element is the chemical potential $\mueid$ of non-interacting electrons
which determines the Fermi occupation factors $g_i$ and $g_k$ in the electron density (Eq. \ref{fermi}). This can be determined
in several different ways. A very common practice \cite{liberman, wilson, piron3, xu}
is to use a local charge neutrality condition. For example
\begin{equation}
  Z=\int_0^R n_e(\mueid,r)\,d^3r,
  \label{neutrality}
\end{equation}
where the $\mueid$ dependence of $n_e(r)$ is shown explicitly, ensures neutrality within the volume of the ion sphere.
This relation uniquely determines $\mueid$. The free electron density that corresponds to $V_{Ne}^{\rm eff}(r \rightarrow \infty)=0$ 
then follows from
\begin{equation}
  n_e^0=\frac{\sqrt{2}}{\pi^2 \beta^{3/2}} \int \frac{x^{1/2}\,dx}{e^{x-\beta\mueid} + 1},
\end{equation}
which also determines the ion charge $Z^*$ (Eq. \ref{zstar}).  
It is well known that there is no rigorous definition of the ion charge in AA models \cite{piron3, sterne}.
While Eq. (\ref{neutrality}) is intuitively sensible, there are other intuitively reasonable
choices \cite{sterne, chihara91, perrot2, crowley1, rozsnyai}, such as requiring that $Z=N_b + n_e^0/n_I^0$, 
where $N_b$ is the number of electrons bound to the central nucleus 
\begin{equation}
  N_b=2 \int \sum_{i={\rm bound}} g_i|\psi_i(r)|^2 \,d^3r
\end{equation}
(see Eq. \ref{e_density}). 
This is equivalent to requiring that the ion charge is that of the nucleus minus the number of bound electrons: $Z^*=Z-N_b$.
It has been shown \cite{piron3} that requiring that the AA model gives the same pressure whether obtained from the virial or thermodynamic routes
gives a condition for $\mueid$ that involves the electrostatic potential rather than the electron density:
\begin{equation} 
  \int_0^\infty d^3r \, \Theta(r-R) V_{Ne}^{\rm el}(r)=0,
\end{equation} 
where
\begin{equation} 
  V_{Ne}^{\rm el}= -\frac{Z}{r} +  \int \frac{n_e(r^\prime)-Z^*\Theta(r^\prime-R)}{|\vec r-\vec r^\prime|}\,d^3r^\prime.
\end{equation} 
This result is obtained  by minimizing the AA free energy with respect to $n_e^0$. Paradoxically, this
more rigorous approach to determine $\mueid$ gives anomalously low values of $Z^*$ at low temperatures \cite{piron3}
, where it fares worse than models using a local neutrality condition such as Eq. (\ref{neutrality}). 
It appears that this more rigorous
AA model exposes a detrimental aspect of neglecting the correlations in the surrounding plasma, a feature that must be
partly compensated in models that use an ad hoc local neutrality condition.

To summarize, the AA model is a DFT-based, spherically symmetric, self-consistent field atom model embedded in a
plasma. The equations are closed with two additional inputs:
1) an external ion distribution function which defines a cavity in a homogeneous (uncorrelated) plasma,
and 2) a criterion to determine the non-interacting (field-free) part of the chemical potential of the electrons.  The latter cannot be
uniquely defined and various sensible choices give different results for the electronic structure of the central ion.
The former is a simple and convenient approximation that can be improved upon.  In fact, the absence of plasma  correlations
in the AA model is the most significant piece of physics that is missing in AA models.  We now expand the AA model
to include interactions within the surrounding plasma and with the central ion while preserving the assumption of
spherical symmetry.

\section{An average atom model with plasma correlations}
\label{aa_tcp}
\begin{figure}
  \includegraphics[scale=.655, trim= 10mm 50mm 0mm 15mm, clip]{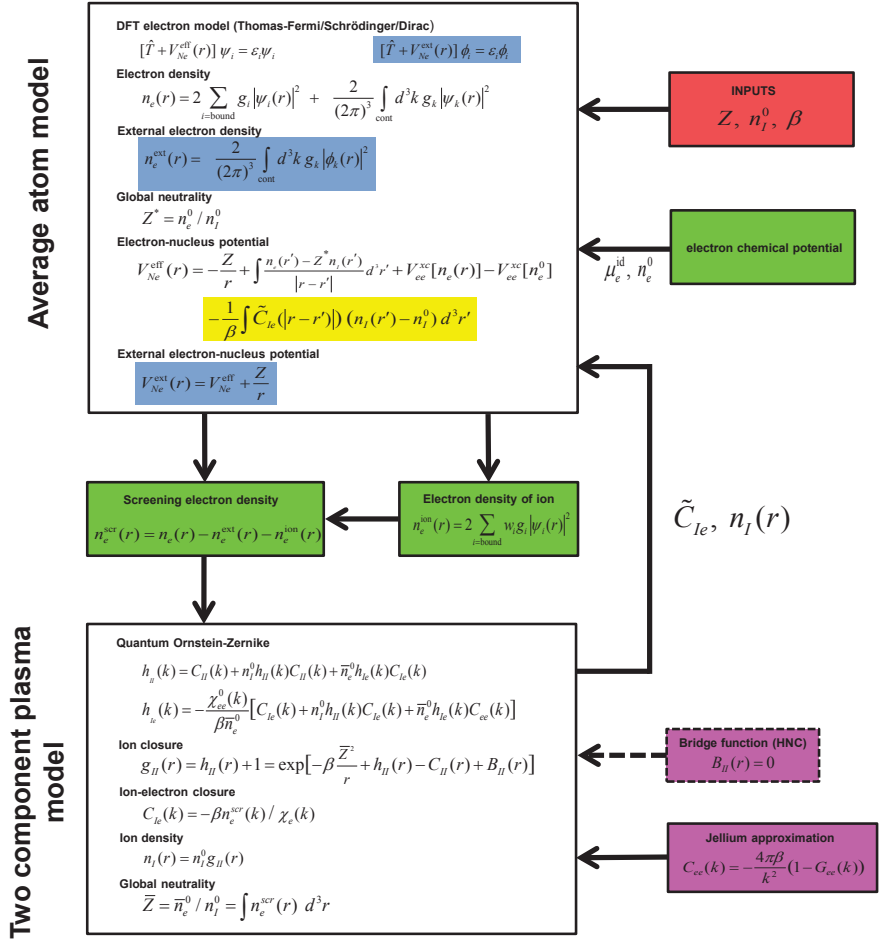}
  \caption{Structure and key equations of the coupled average atom and two-component plasma model. The red box gives the 
           input parameters to the model.  Green boxes show elements where other choices are possible. 
           Approximations are shown within purple boxes, with the dotted box indicating an approximation that can readily be improved. 
           Arrows show the flow of information between the elements of the model.  See text for details.}
  \label{fig:aa_tcp_diag}
\end{figure}

The importance of the correlations in the surrounding plasma was recognized decades ago and  several models
have been developed to account for them \cite{scaalp, chihara91, perrot2, xu}. They affect the
AA model shown in Fig. \ref{fig:aa_diag} in two ways. First the effective nucleus-electron potential now includes
interactions between the electrons and the surrounding ions: 
\begin{align}
  V_{Ne}^{\rm eff}(r) &= -\frac{Z}{r} + \int \frac{n_e(r^\prime)-Z^*n_I(r^\prime)}{|\vec r-\vec r^\prime|}\,d^3r^\prime + \Vxc[n_e(r)] - \Vxc[n_e^0]  \nonumber \\
                      & \quad -\frac{1}{\beta} \int {\tilde C}_{Ie}(|\vec r-\vec r^\prime|)(n_I(r^\prime)-n_I^0)\,d^3r^\prime,\hfill
  \label{VNe_tcp}
\end{align}
where ${\tilde C}_{Ie}$ is the non-Coulombic part of the direct ion-electron correlation function $C_{Ie}(r)$, which is defined by \cite{starrett2}
\begin{equation}
  C_{Ie}(r)=\beta Z^*/r + {\tilde C}_{Ie}(r). 
  \label{Ctilde}
\end{equation}
This new term is highlighted in yellow in Fig. \ref{fig:aa_tcp_diag}. The second modification is that the ion density is described
in terms of the ion-ion radial distribution function $g_{II}(r)$:
\begin{equation}
  n_I(r)=n_I^0 g_{II}(r).
\end{equation}
Whereas in the AA model $g_{II}(r)$ is chosen as a simple step function, here it is to be calculated on the basis of the 
Coulomb interactions in the plasma. As can be seen by comparing Figs. \ref{fig:aa_diag} and \ref{fig:aa_tcp_diag}, the structure of the 
AA model remains unchanged except for the new contribution to $V_{Ne}^{\rm eff}(r)$. The inputs to the model ($Z$, $n_I^0$, $\beta$)
are the same and a criterion for $\mueid$ must still be provided.  In addition, closure of this modified
AA model also requires ${\tilde C}_{Ie}(r)$ and $n_I(r)$ which can be calculated by coupling the AA model with a model of the plasma based on
the Ornstein-Zernike integral equations theory of fluids \cite{starrett2}. The calculation of these two quantities increases the 
complexity of the problem considerably (Fig. \ref{fig:aa_tcp_diag}).

\subsection{Ornstein-Zernike integral equations theory of fluids and the two-component plasma model}

Prior to the advances in computing power that allowed the direct simulation of
classical fluids (Monte Carlo and Molecular Dynamics methods), the integral equations  theory of fluids \cite{hansen1,caccamo}
offered the most realistic description of an interacting fluid at a reasonable computing cost. It is the theory of choice to
develop an AA model with correlations \cite{chihara91, anta, perrot2, peyrusse, xu, starrett1, starrett2, ofer}.

The structure and thermodynamics of
a fluid system of one type of classical point particles interacting via a pair potential is fully determined by its density $n$,  
temperature $\beta$ and the pair potential $V(r)$. The most widely used integral equations theories of fluids are based
on the Ornstein-Zernike equation 
\begin{equation}
  h(r)=C(r) + n\int C(|\vec r - \vec r^\prime|)h(r)\,d^3r^\prime.
  \label{oz_r}
\end{equation}
Expressed in Fourier space, Eq. (\ref{oz_r}) takes a simple form
\begin{equation}
  h(k)=C(k) + nC(k)h(k),
  \label{oz_k}
\end{equation}
where $F(k)$ denotes the Fourier transform of a function $F(r)$.  The functions $h(r)$ and $C(r)$ are related to the pair potential by a closure relation 
\begin{equation}
  g(r)=h(r)+1=\exp \left(-\beta V(r) + h(r)-C(r)+B(r) \right).
  \label{oz_closure}
\end{equation}
In these equations,
$g(r)$ is the pair distribution function, $h(r)$ is the pair correlation function, and $C(r)$ is the direct correlation function which
is effectively defined by Eq. (\ref{oz_r}).
Closing these two equations requires the knowledge of the bridge function $B(r)$ which accounts for $n$-body correlations beyond 
pair correlations. Various approximations to $B(r)$ are available \cite{caccamo}.  Note that these relations are exact if the bridge function is known.
Once the structure of the fluid (e.g. $g(r)$) is known, various quantities of interest, such as the thermodynamics can be
calculated \cite{hansen1}.

The above fluid equations can be readily generalized to mixtures of classical particles given the pair potentials $V_{ij}(r)$ between
particles of type $i$ and $j$ \cite{hansen1,caccamo}. Chihara \cite{chihara85} has shown that the Ornstein-Zernike equation can be further generalized for a plasma of classical ions
of density $n_I^0$ and quantum electrons of density $\nebar$
\begin{eqnarray}
  h_{II}(k)= & C_{II}(k) + n_I^0 h_{II}(k)C_{II}(k) + \nebar h_{Ie}(k)C_{Ie}(k) \\
  h_{Ie}(k)= & {\frac{-\chi_{ee}^0(k)}{\beta \nebar}} \left[C_{Ie}(k) + n_I^0 h_{II}(k)C_{Ie}(k) + \nebar h_{Ie}(k)C_{ee}(k) \right]
  \label{oz}
\end{eqnarray}
where $\chi_{ee}^0(k)$ is the Lindhard response function of the non-interacting quantum electron gas.
The quantum nature of the electrons is embodied in the factor $-\chi_{ee}^0(k)/{\beta \nebar}$. In the limit of a classical
electron gas, this factor becomes unity and the classical Ornstein-Zernike equations are recovered.  
The ion closure relation has the same form as in the one component case
\begin{equation}
  g_{II}(r)=h_{II}(r)+1=\exp \left(-\beta \frac{{\bar Z}^2}{r} + h_{II}(r)-C_{II}(r)+B_{II}(r) \right)
\end{equation}
where the pair interaction is a pure Coulomb potential between ions of charge $\Zbar$.\footnote{Note that this is different from the AA ion
charge $Z^*$, which is further discussed in \S \ref{sec:plasma_corr}.}   Global neutrality requires that 
\begin{equation}
  \Zbar= \nebar/n_I^0. 
  \label{zbar}
\end{equation}
The last four equations contain seven unknowns: ${\bar Z}$, $h_{II}$, $h_{Ie}$, $C_{II}$, $C_{Ie}$,
$C_{ee}$ and $B_{II}$. The closing of this system of equations will allow a solution that provides ${\tilde C}_{Ie}$ (from Eq. \ref{Ctilde})
and $n_I(r)=n_I^0 g_{II}(r)$ that are needed to close the AA model with ion correlations (Fig. \ref{fig:aa_tcp_diag}). The full set 
of equations will define a model that we will call a ``two-component plasma'' (TCP) model, 

To close the equations of the TCP model, we start by introducing two approximations (purple boxes in Fig. \ref{fig:aa_tcp_diag}). First, the
bridge function is assumed to be $B_{II}(r)=0$. This is the well-known hyper-netted chain (HNC) approximation to the Ornstein-Zernike integral equations
theory of fluid (dashed purple box in Fig. \ref{fig:aa_tcp_diag}).
A better approximation is to use the bridge function of a similar interacting system that is more easily calculated than
that of a general TCP.
The bridge functions of the hard sphere fluid \cite{rosenfeld79,labik}, of the one-component plasma model \cite{iyetomi92}, or for Yukawa
systems \cite{daughton00}, or bridge functions optimized with a variational approach \cite{rosenfeld}
have all been used to model plasmas with
the integral equations theory of fluids.  For simplicity, and with good accuracy for the systems considered here (see \S \ref{results}), we will use
the HNC approximation.  The second approximation is to assume that the direct electron-electron correlation function $C_{ee}$ 
is given by 
\begin{equation}
  C_{ee}(k)=\frac{-4\pi\beta}{k^2} \left( 1-G_{ee}(k,\beta,\nebar) \right)
  \label{cee}
\end{equation}
where $G_{ee}(k,\beta,\nebar)$ is the local field correction of the jellium model.
The jellium approximation (Eq. \ref{cee}), to which there appears to be no reasonable alternative at this time, implies that the $e$--$e$ correlations are 
decoupled from the ions. This is reasonable as $C_{ee}$ represent the correlations between unbound electrons. As long as the ion-electron
correlations are not too strong, Eq. (\ref{cee}) is a good approximation.  On the other hand, it would be a very poor
approximation to describe bound electrons which are intrinsically very strongly correlated with the ions.

One additional equation is needed to close the TCP model, which is to be provided by the AA model (downward arrows in Fig. \ref{fig:aa_tcp_diag}). 
This closure will also determine the charge of the ion $\Zbar$ and thus the direct ion-ion potential in the TCP 
model.  This closure is not unique, however \cite{starrett1, starrett2, chihara91, perrot2}, and
is illustrated with the green boxes in Fig. \ref{fig:aa_tcp_diag}. We found that the most satisfactory approach is to use the solution
of the AA model to calculate the electron density that screens an individual ion in the plasma.  The AA model is naturally suited for this purpose
as it is designed to calculate the properties of a single ion in a plasma. This screening electron density is defined in two steps.

In the first step, the
electron density profile $n_e(r)$ obtained from the AA model is separated into a contribution that is assigned to a single nucleus
(defining a ``pseudo-atom'' \cite{ziman67, dagens72, perrot2}) and an external contribution, due to the electrons in the surrounding plasma.  The separation 
is effected by solving the AA model a second time after removing the central nucleus 
from the effective electron-nucleus potential $V_{Ne}^{\rm eff}(r)$, keeping everything else fixed.  
The resulting electron density is thus computed only from the field of the
surrounding plasma $V_{Ne}^{\rm ext}(r)$ and defines the external electron density $\neext(r)$. The AA equations corresponding 
to this external system are shown in blue boxes in Fig. \ref{fig:aa_tcp_diag}. Note that in the absence of the central nucleus, the potential
$V_{Ne}^{\rm ext}(r)$ is repulsive almost everywhere and does not support bound states. This pseudo-atom electron density is 
$n_e^{\rm PA}(r)=n_e(r)-\neext(r)$. 

The second step consists of extracting from $n_e^{\rm PA}(r)$ the part that will define
(along with a central nucleus) an ion, and the remainder which is to be assigned to the electron fluid of the plasma. The most straightforward
separation is to assign the bound electrons to the ion and the continuum electrons to the TCP electron fluid. This
definition of an ion for the TCP model (green box in Fig. \ref{fig:aa_tcp_diag}) is by no means unique. For instance, the 
density in the neighborhood of the nucleus associated with continuum resonances of low energy is qualitatively similar to
the electron density of a weakly bound state. Thus, it would be reasonable to count those resonant electrons as part of the ion.
Such an approach was suggested by Chihara \cite{chihara91} but is not satisfactory in practice as
it requires additional arbitrary criteria to determine which resonances should be considered part of the ion and to separate the electron density
associated with a resonance from the background continuum electrons. 

This simple definition of the ion electron cloud is not without 
drawbacks. Most flagrant is that the electron density of the bound states is a discontinuous function of $n_I^0$ when a
state crosses into the continuum of positive energies, i.e. when it is pressure-ionized,  a phenomenon of great importance in studies of WDM. As a 
consequence, every quantity appearing in the model becomes discontinuous with density. This can be avoided by 
applying occupation factors $0 \le w_i(r) \le 1$ in the sum 
\begin{equation}
  \neion(r)=2\sum_{i={\rm bound}} w_i(r) g_i |\psi_i(r)|^2
  \label{neion}
\end{equation}
which are calculated from a simple model for the broadening of the bound energy levels \cite{starrett2}. As a weakly bound, broadened level starts
to cross into the continuum of positive energies, $w_i(r)$ decreases below unity and vanishes once $\epsilon_i=0$. Furthermore, $w_i(r)$ includes a
smooth radial cutoff that damps the tails of weakly bound states that extend far from the central nucleus \cite{perrot2, starrett2}. Physically,
this accounts for the fact that the bound electron density that is located far from the nucleus cannot be assigned to the central ion in 
the presence of the other neighboring ions.  This cutoff also resolves a numerical difficulty with the spatial integration of long-ranged functions.
With these elements it is now possible to define the screening electron density
\begin{equation}
  n_e^{\rm scr}(r)=n_e(r)-\neext(r)-\neion(r).
  \label{nescr}
\end{equation}
This is the density of electrons that are associated with the central  nucleus but are ``free'', i.e. not part of the central ion. These electrons respond
to the attractive field of the ion, screening the ion-ion interaction. From the quantum OZ equations, it can be shown \cite{anta, chihara91, starrett2}
that the response of the electron fluid gives the screening density in terms of direct correlation functions
\begin{equation}
  n_e^{\rm scr}(k) = -\chi_{ee}(k) C_{Ie}(k)/\beta
  \label{coupling1}
\end{equation}
where
\begin{equation}
  \chi_{ee}(k) = \frac{\chi_{ee}^0(k)}{1+\chi_{ee}^0(k) C_{ee}(k)/\beta}
\end{equation}
is the response function of the correlated quantum electron fluid \cite{ashcroft78}.  Equation (\ref{coupling1})
has the form of a linear response of the electron fluid to
an external pseudo-potential $-C_{Ie}/\beta$.  However, in this case the response is highly non-linear since $n_e^{\rm scr}(r) $ is calculated
from the solution of the Schr\"odinger equation rather than linear response theory. Inverting Eq. (\ref{coupling1}) gives
\begin{equation}
  C_{Ie}(k)=-\beta n_e^{\rm scr}(k)/\chi_e(k)
  \label{Cie}
\end{equation}
which is an exact relation within the TCP model. Furthermore, global neutrality of the TCP plasma requires that the ion charge ${\bar Z}$ be 
related to the screening density by
\begin{equation}
  {\bar Z}=\nebar/n_I^0 = \int n_e^{\rm scr}(r) \, d^3r.
\end{equation}
By taking $n_e^{\rm scr}$ defined from the AA model (Eq. \ref{nescr}), we now 
have the final equation required to close the TCP model, which provides ${\tilde C}_{Ie}$ (Eq. \ref{Ctilde}) and
$n_I(r)$ that in turn  close the AA model (Fig. \ref{fig:aa_tcp_diag}). 
The two sets of equations are solved iteratively between the AA and TCP models. 

Thus, the coupling between the AA and the TCP models is provided by defining a screening density from a solution of the
Schr\"odinger equation for the $Z$ electrons belonging to the central nucleus in the AA model. This particular choice is
not unique (hence the green box in Fig. \ref{fig:aa_tcp_diag}). In the same spirit, the Quantum Hypernetted-Chain (QHNC) model of Chihara \cite{chihara91, anta},
 couples the AA and TCP models in terms of the pair correlation function
\begin{equation}
  n_e^0 h_{Ie}(r)=n_e^{\rm f}(r)-n_e^0
\end{equation}
where $n_e^{\rm f}(r)$ is the density of continuum electrons (second term in Eq. \ref{e_density}). Like Eq. (\ref{Cie}), this choice for the coupling
relates a quantity associated with the density of free electrons obtained in
the AA model ($n_e^{\rm f}$ instead of $\nescr$) to a correlation function in the TCP model ($h_{Ie}$ 
instead of $C_{Ie}$). The QHNC closure has the drawback
that $n_e^0={\nebar}$, i.e. ${\bar Z}=Z^*$, which overly constrains the model, as we will see below.

The electron chemical potential (green box in Fig. \ref{fig:aa_tcp_diag}) is given as in the AA model (Eq. \ref{neutrality}), 
with a step function $g_{II}(r)$.  This is inconsistent with the pair distribution function obtained in the TCP model but is numerically
advantageous and, given the ambiguous nature of the local neutrality condition, is very reasonable.  
Other choices include imposing charge neutrality as in Eq. (\ref{neutrality}) but over a correlation sphere of radius $R_c>R$ \cite{peyrusse},
or $Z=N_b + n_e^0/n_I^0$ as in several AA models.

This completes the formulation of the self-consistent model for an average atom embedded in a plasma of correlated ions. This AA+TCP model assumes
spherical symmetry and
consists of an AA model coupled to a TCP model based on the Ornstein-Zernike integral equations theory of interacting fluids. A screening electron density is
calculated within the AA model to obtain an ion-electron correlation function for the TCP model, which in turns provide correlation functions
for the AA atom model. The model has no adjustable parameters and only needs the nuclear charge, ion density and temperature as inputs.
It formally recovers the classical one-component plasma model (OCP) \cite{ichimaru_rmp} and the screened one-component plasma model (SOCP) \cite{chabrier}.
The solution of the closed system of equations gives all the pair correlation functions between ions and electrons, the effective ion-ion 
and nucleus-electron pair potentials, the bound and continuum wave functions and eigenvalues, the ion charge, and the electron densities.
These quantities, some of which are directly amenable to experimental measurements, form the basis upon which
WDM properties of interest, such as the equation of state, conductivities, and opacities can be calculated.

\subsection{The role of plasma correlations}
\label{sec:plasma_corr}

\begin{figure}
  \centering
  \includegraphics[scale=.75, trim = 10mm 55mm 100mm 120mm, clip]{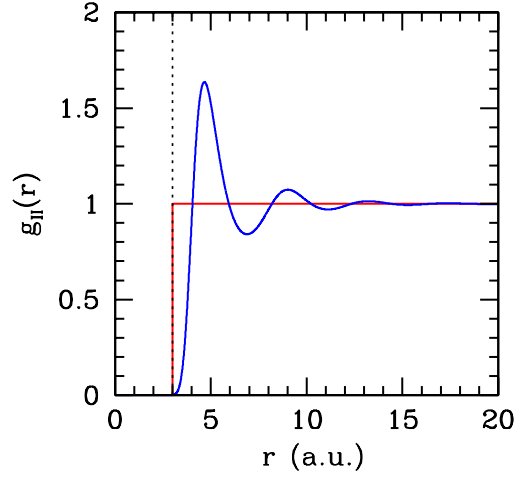}
  \caption{Pair distribution functions of the JVM model (red) and the AA+TCP model (blue) for Al at $T=1\,$eV and $\rho=2.7\,$g/cm$^3$.
           At this density, the ion sphere radius is $R=2.99\,$a.u., which is indicated by a dotted line.}
  \label{fig:g_II}
\end{figure}

\begin{figure}
  \centering
  \includegraphics[scale=.75, trim= 10mm 57mm 100mm 120mm, clip ]{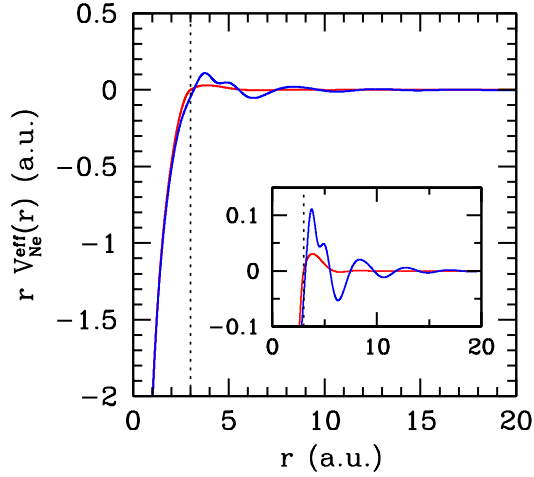}
  \caption{Effective nucleus-electron potential for the JVM (red) and the AA+TCP model (blue), from Eqns. (\ref{VNe}) and (\ref{VNe_tcp}), respectively,
           for the same conditions as in Fig. \ref{fig:g_II}.
           The inset magnifies the behavior at larger $r$.  The product $rV_{Ne}^{\rm eff}(r)$ is plotted for clarity and the dotted
           line indicates the ion sphere radius $R=2.99\,$a.u..}
  \label{fig:v_ne}
\end{figure}

\begin{figure}
  \centering
  \includegraphics[scale=.75, trim = 10mm 57mm 100mm 120mm, clip]{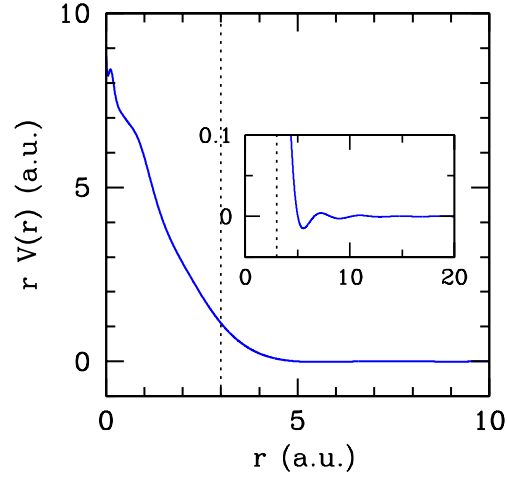}
  \caption{Effective ion-ion potential in the the AA+TCP model, given by Eqn.(\ref{v_ii}) for the same case as in Fig. \ref{fig:g_II}. 
           The inset magnifies the behavior at larger $r$.  The ion sphere radius $R=2.99\,$a.u. is shown by
           the dotted line.  The product $rV(r)$ is plotted for clarity.}
  \label{fig:v_ii}
\end{figure}

\begin{figure}
  \centering
  \includegraphics[scale=.70, trim= 10mm 57mm 100mm 90mm, clip]{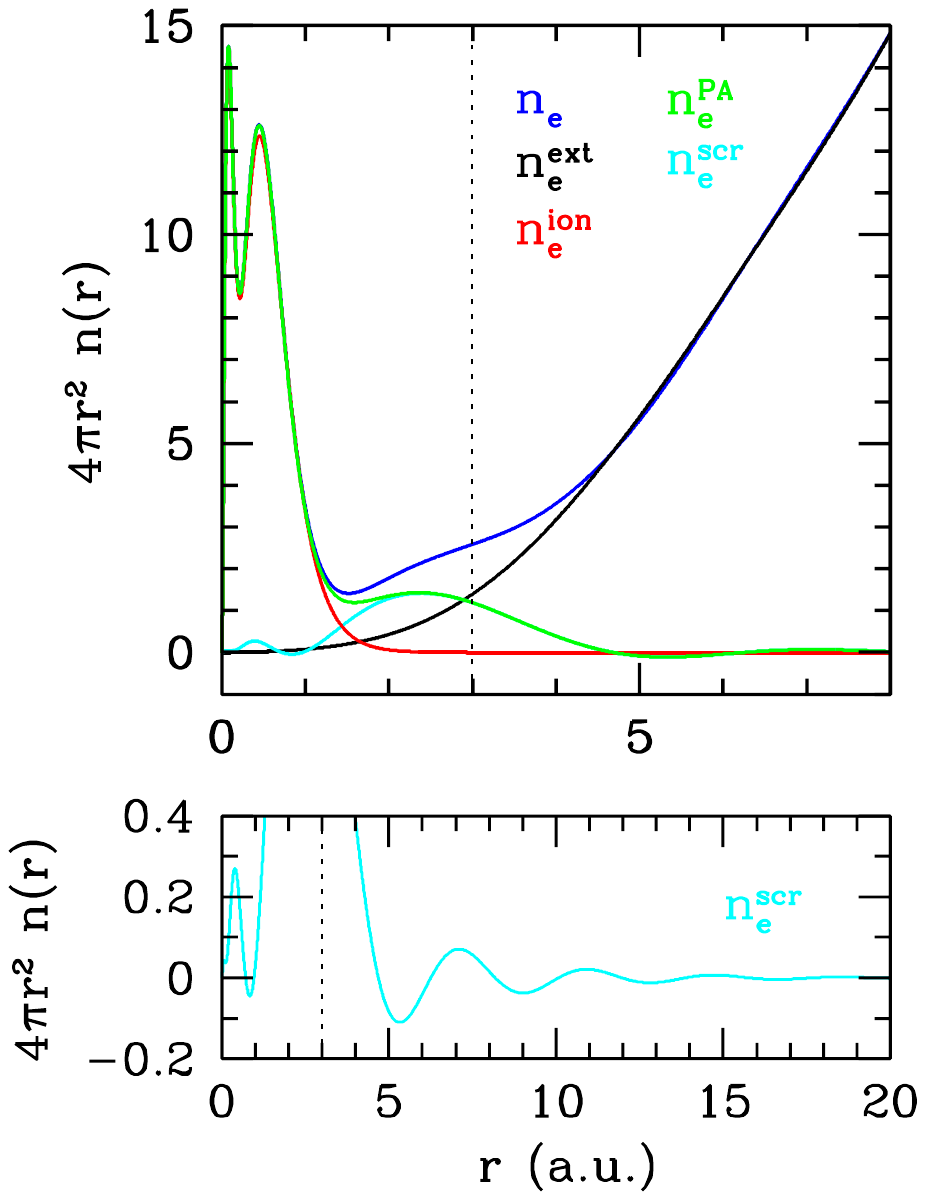}
  \caption{Electron densities for Al at $T=1\,$eV and 2.7$\,$g/cm$^3$. Only the electron densities from the AA+TCP model
           are shown for clarity. The total electron density (Eq. \ref{e_density}) is in blue, and the density from the system external to
           the ion (i.e. without the central nucleus) $\neext$ is in black.  The difference between them is the pseudo-atom
           density $n_e^{\rm PA}= n_e - \neext$, which represent the $Z$ electrons, bound and unbound, associated with the
           central nucleus (green). The density of electrons which, along with the central nucleus, define an ion is $\neion$ (Eq.
           \ref{neion}, red).  Finally, the screening density $\nescr=\nepa-\neion$ is shown in cyan. The lower panel, which is on 
           different scale, shows the smaller structure of $\nescr$. 
           In the upper panel, the $n_e$ and $\neext$ curves diverge as $r^2$ at large $r$ because both densities become constant and
           the figure shows the densities multiplied by $4\pi r^2$. The vertical dotted line shows the ion sphere radius $R=2.99\,$a.u.}
  \label{fig:ne}
\end{figure}

The role of ion-ion correlations in the AA+TCP model is revealed by comparing some key model quantities calculated
with an imposed, step function pair distribution function $g_{II}(r)$ (Eq. \ref{ni}) as in a typical AA model and with the full self-consistent $g_{II}(r)$
from the AA+TCP model. For clarity, we will call the former the jellium vacancy model (JVM).\footnote{This is the same model as the JVM of \cite{piron3}
and similar to the JVM of \cite{chihara91}.} These results are shown in Figs. \ref{fig:g_II} -- \ref{fig:ne} for the case of Al at $T=1\,$eV 
(0.0367$\,$Ha) and $\rho=2.7\,$g/cm$^3$
(solid density). Under these conditions, the ions are strongly correlated and $g_{II}(r)$ shows much structure 
which is only crudely approximated by the step function ion distribution of the JVM (Fig. \ref{fig:g_II}). 

The effect of the ion correlations on the effective nucleus-electron potential $V_{Ne}^{\rm eff}(r)$ that enters the Schr\"odinger
equation of the AA model is shown in Fig. \ref{fig:v_ne}. At this density, the 1$s^2$, 2$s^2$ and 2$p^6$ electrons of Al are tightly
bound (our model predicts binding energies of -54.6, -3.4 and -2.0$\,$Ha, respectively, measured from the bottom of the valence band) 
and confined to $r \wig< 2\,$a.u. (see below). Thus the core electrons are
not sensitive to the difference in $g_{II}(r)$ and the two potentials are 
identical in the core. Both show a small positive maximum
at $r\sim R$ as the central ``ion hole'' (Fig. \ref{fig:g_II}) amounts to a repulsive contribution to the effective nucleus-electron 
potential. Differences, which can be as large as 0.02$\,$Ha (0.5$\,$eV), are mainly 
confined to $R \wig< r \wig< 3R$, as is expected from the difference in the ion-ion distribution function (Fig. \ref{fig:g_II}).
At this relatively low temperature, $V_{Ne}^{\rm eff}(r)$ has Friedel oscillations at large $r$ superimposed
on the effect of decaying oscillations in $g_{II}(r)$.  Both go away at higher $T$ \cite{starrett2, saumon_wdm}.

In the TCP model, an effective ion-ion pair potential that takes into account the electron screening can be defined \cite{starrett2, anta, chihara91}
\begin{equation}
  \beta V(k)= \beta \frac{4\pi \Zbar^2}{k^2} - n_e^{\rm scr}(k) C_{Ie}(k).
  \label{v_ii}
\end{equation}
The first term is the repulsive Coulomb potential between two point ions of charge $\Zbar$ and the second term is the screening 
(mainly attractive) potential from the surrounding electrons that
ensures that $V(r)$ is short-ranged.  This effective ion-ion potential is plotted in Fig. \ref{fig:v_ii} as $rV(r)$. The 
ion charge is $\Zbar=3$ and a pure Coulomb potential would appear as a flat line at $rV(r)=9$. The ion-ion potential reaches this limit as $r \rightarrow 0$.
Strong screening of the ion charge is evident as $rV(r)$ decreases rapidly to very small values within $\wig< 2R$. The potential shows two small bumps
at $r\sim 0.07$ and 0.85$\,$a.u. that are associated with minima in $n_e^{\rm scr}(r)$ (see below).  This is consistent with Poisson's equation for 
the electrostatic contribution of $\nescr$ to the ion-ion potential.
Small oscillations outside the ion sphere (inset) are caused by the structure in $g_{II}(r)$.

The various electron densities that appear in the model, $n_e(r)$, $\neext(r)$, $n_e^{\rm PA}(r)$, $\neion(r)$ and $\nescr(r)$ are shown in
Fig. \ref{fig:ne}.  At the scale of this figure, these electron densities are nearly identical when computed with either the JVM or the AA+TCP model.
Only the latter are shown for clarity. The total electron density $n_e(r)$ (Eq. \ref{e_density}, blue curve)
shows two peaks within 1$\,$a.u. that correspond to the 1$s^2$ and the 2$s^2$2$p^6$ bound electrons.  The continuum electrons that are attracted to the 
nucleus and its bound electrons cause an excess of density near the ion sphere radius.
At large $r$, $n_e(r) \rightarrow n_e^0$ and
$4\pi r^2 n_e(r)$ diverges as $r^2$.  The electron density associated with the plasma external to the central nucleus, $\neext(r)$ (black curve) is
similar to $n_e(r)$ except that it does not support bound states and has no build up of screening charge.  The difference between these two
electron densities defines the electron cloud associated with the central nucleus, i.e. a pseudo-atom $\nepa(r)=n_e(r)-\neext(r)$ (green curve), which
is essentially confined within $\sim 2R$. The ion electron cloud ($\neion(r)$) is defined as the density that arises from 
the bound states (Eq. \ref{neion};
red curve) which shows peaks associated with the K and L shells.  Finally, the electron density that couples the AA and TCP model (Fig. \ref{fig:aa_tcp_diag}
and Eq. \ref{Cie}) is the screening density $\nescr(r)=\nepa(r)-\neion(r)$ (cyan curve). It forms a broad peak outside of the core states
that decays rapidly beyond the ion sphere radius.  This results from the attractive force of the net positive charge of the ion (nucleus + bound states),
the repulsion from the bound electrons and the orthogonality of the continuum and bound states. Details of $\nescr(r)$ are shown in the lower 
panel of Fig. \ref{fig:ne}.

It turns out that by running the electron densities obtained with the JVM through a single pass of the TCP model (without requiring the self-consistency,
see Fig. \ref{fig:aa_tcp_diag}) gives distribution functions that are good approximations to the fully converged solutions \cite{starrett2}. This realization
has been a key element of the successful numerical implementation of the model as it substantially improves the convergence.  It also implies 
that approximate but fairly good $g_{II}(r)$ for
the AA+TCP model can be calculated at very little computational cost above that of the JVM model.

\begin{figure}
  \centering
  \includegraphics[scale=.7, trim= 10mm 57mm 100mm 120mm, clip]{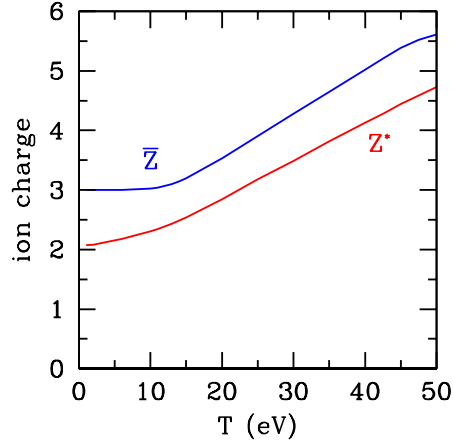}
  \caption{Ion charge as a function of temperature for solid density Al (2.7$\,$g/cm$^3$). The ion charge from
           the coupled average atom $Z^*$ and the two-component plasma models ${\bar Z}$ are shown. The roles
           of $\Zbar$ and $Z^*$ are illustrated in Fig. \ref{fig:aa_tcp_diag}.}
  \label{fig:Z}
\end{figure}

As pointed out above, the ion charge in the AA part of the model, $Z^*$, and in the TCP part of
the model, ${\bar Z}$ are two distinct quantities, as are the corresponding free electron densities, $n_e^0$ and $\nebar$. 
It should be clear from Fig. \ref{fig:aa_tcp_diag} that we are coupling two different models
and that mathematically it is not required that these ion charges be equal. In fact, they are quite different, as can
be seen in Fig. \ref{fig:Z}. As expected, the ion charge
increases steadily with temperature. The counterintuitive distinction
between $\Zbar$ and $Z^*$ can be understood physically as follows. In the AA model, $n_e^0$ is the
density of electrons in the field-free region of space ($V_{Ne}^{\rm eff}=0$), far from the central nucleus, which has a clear
meaning within a model with a single central nucleus embedded in a spherically averaged plasma.  In the TCP model, on the other hand, $\nebar$ represents 
the density 
of the inhomogeneous electron fluid (made up of those electrons that are not included in the ions), averaged over all space. This corresponds more closely 
to the concept of the electron density in a real dense plasma, where there is no field-free region as an electron always finds 
itself in the field of some nearby ion(s). Thus, the free electron density in the AA model ($n_e^0$) is a rather artificial 
construct that arises from the simplifying concept that
the electronic structure around a central nucleus in a smeared out plasma is a good approximation for the 
``average'' electronic structure in a correlated, multi-center plasma. Within the AA+TCP model, the physical electron
density (to be used in a conductivity calculation, for example)  is $\nebar$, with corresponding ion charge $\Zbar$.  
This is confirmed in Fig. \ref{fig:Z} where $\Zbar \rightarrow 3$
at low $T$, which is the number of valence electrons of Al under normal conditions.  On the other hand, $Z^* \sim 2$ at low $T$, which would be a
poor estimate of the valence electron density in normal aluminum, should it be (wrongly) interpreted as such. 

\subsection{Appraisal of the model}
\label{pros_cons}

Given the theoretical development and numerical solutions of the model outlined above, it is 
possible to ascertain its advantages and limitations at a semi-quantitative level. 
There are several advantages over computer simulations of WDM. Most immediate is the substantial economy of
computer time to converge to a solution.  Depending on the temperature and density,
the solution of the AA+TCP model is typically 2 to 3 orders of magnitude faster than a QMD simulation. Computer simulations
are inherently subject to statistical noise (fluctuations) due to the finite number of particles considered. 
The finite size of the simulation box limits spatial sampling of the system, such as the radial extent over
which the pair distribution function  can be evaluated (Figs. \ref{fig:Al_g} -- \ref{fig:Fe_g}).  
On the other hand, the AA+TCP model consists of coupled 
algebraic, differential and integral equations whose solution is smooth within the numerical accuracy
of the algorithms. 
The equations are solved over a comparatively large computational volume of typically 10--20 times the
ion sphere radius, which effectively corresponds to an infinite system.  It also
treats all of the electrons explicitly and on the same footing. There is no pseudo-potential or concern about their
transferability. The model gives good solutions over a range of temperatures and densities that is
much broader than the typical range of applicability of any one {\it ab initio} method. It is also worth noting
that a version of the model where the Schr\"odinger equation is replaced by the semi-classical Thomas-Fermi 
model of the electrons gives viable results even for strongly coupled systems \cite{starrett2}, contrary to the conclusions 
of an earlier effort to
develop a Thomas-Fermi AA model with ion-ion correlations \cite{ofer}. Furthermore, the quantum mechanical
version recovers the Thomas-Fermi results in the high-density, high-temperature limit \cite{starrett2}. This is a valuable internal check
on the physics and numerical implementation of the model.

While these are significant advantages over {\it ab initio} simulations of WDM, the model does not fare as well in other
aspects. The substantial savings of computer time come at the cost of more approximate physics.
In particular, the model only considers pair interactions and is restricted to spherical symmetry. It cannot
account for chemical bonding or angular forces. Its range of validity has been demonstrated to be quite broad
in terms of $T$, $\rho$ and $Z$, which shows promise for its usefulness to many potential applications, but 
its limits have not yet been established. It is worth pointing out that the AA model, like QMD, is based on the finite temperature
DFT formalism, and thus both methods share the limitations associated with this approach (such as the
well-known underestimation of the electronic band gap, approximate energies for the bound states, the
fictitious nature of the unoccupied states, etc).

Perhaps of more concern are the elements of the model that readily admit other possible choices (green boxes in Fig. \ref{fig:aa_tcp_diag}). 
This introduces a level of arbitrariness in the model whose effect on the solution is not negligible but has not been quantified.
Our choices have been guided by earlier work, our understanding of the AA and TCP models, and the pragmatic need for a stable numerical
solution of the system of equations. Nonetheless, the choices adopted are reasonable and their validity can be established by assessing 
the accuracy of the results.

\section{Results: Pair distribution functions}
\label{results}

A converged solution of the system of equations that define the model consists of the electronic wave functions and energies of bound states,
the continuum wave functions including resonances, the ion-ion and nucleus-electron effective potentials, the average ion charge, and all 
the correlation and distribution functions 
that describe the structure of the two component ion-electron fluid. Of all these quantities, the ion-ion pair distribution 
function $g_{II}(r)$ is the most readily amenable to a comparison with experimental data and computer simulations. 
The experimental measurement of the structure of WDM is in its infancy and vary sparse \cite{saiz, kritcher}
but a number of results from {\it ab initio} simulations are available.
All the calculations shown in this section were performed in the HNC approximation where
$B_{II}(r)=0$.  This approximation is expected to become gradually worse as the ion-ion coupling increases such as at lower temperatures
or higher ionic densities.

\begin{figure}
  \centering
  \includegraphics[scale=0.90, trim= 10mm 60mm 70mm 47mm, clip]{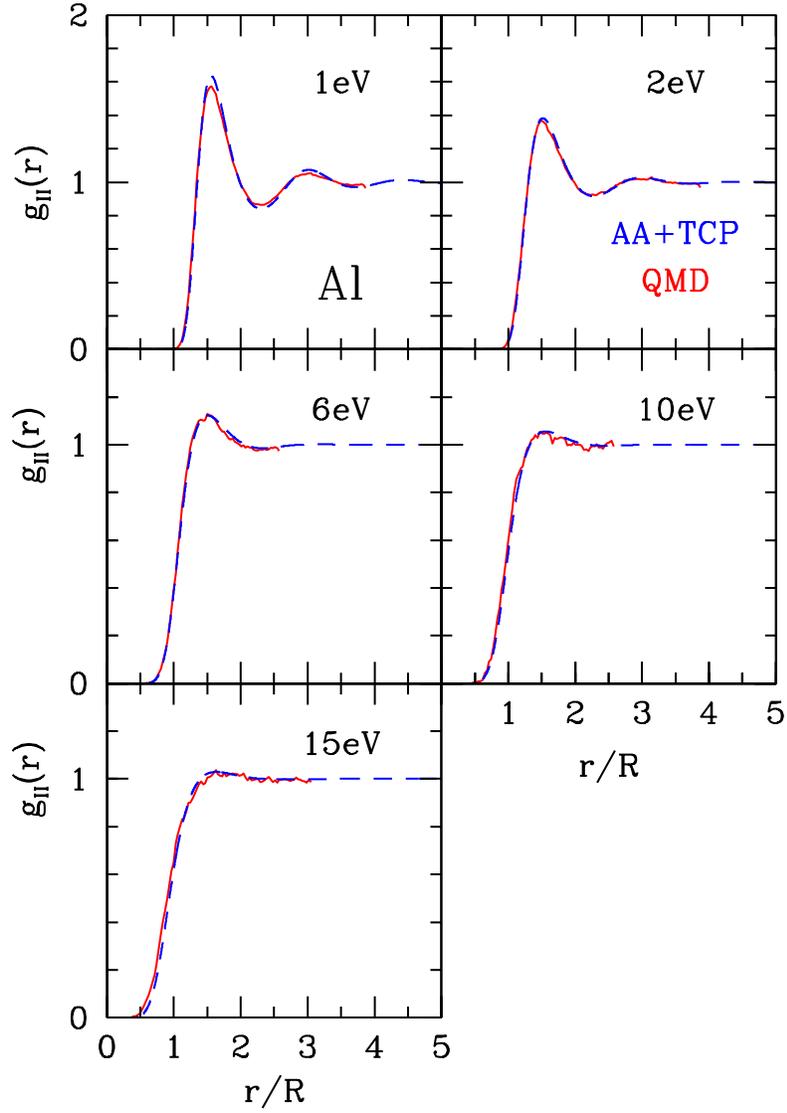}
  \caption{Pair distribution function of aluminum at solid density ($2.7\,$g/cm$^3$) and five temperatures
           ranging from 1 to 15$\,$eV.
           The red curves are quantum molecular dynamics (QMD) simulations \cite{saumon_wdm}.
           The results of the AA+TCP model are shown in blue. $R$ is the ion sphere radius (Eq. \ref{Ris})
           which is 2.99$\,$a.u. in all cases shown. Note that the last three panels are on a different
           vertical scale.}
  \label{fig:Al_g}
\end{figure}

The pair distribution function of aluminum at solid density (2.7$\,$g/cm$^3$) and temperatures from 1 to 15$\,$eV is
compared to the results of QMD simulations \cite{saumon_wdm} in Fig. \ref{fig:Al_g}. The agreement is essentially perfect at $T=2\,$eV and
above and remains very good at 1$\,$eV where deviations likely due to the HNC approximation are discernible. We emphasize that this
excellent agreement is not the result of a fit as there is no adjustable parameter in the AA+TCP model.

\begin{figure}
  \centering
  \includegraphics[scale=0.95, trim= 15mm 60mm 70mm 95mm, clip]{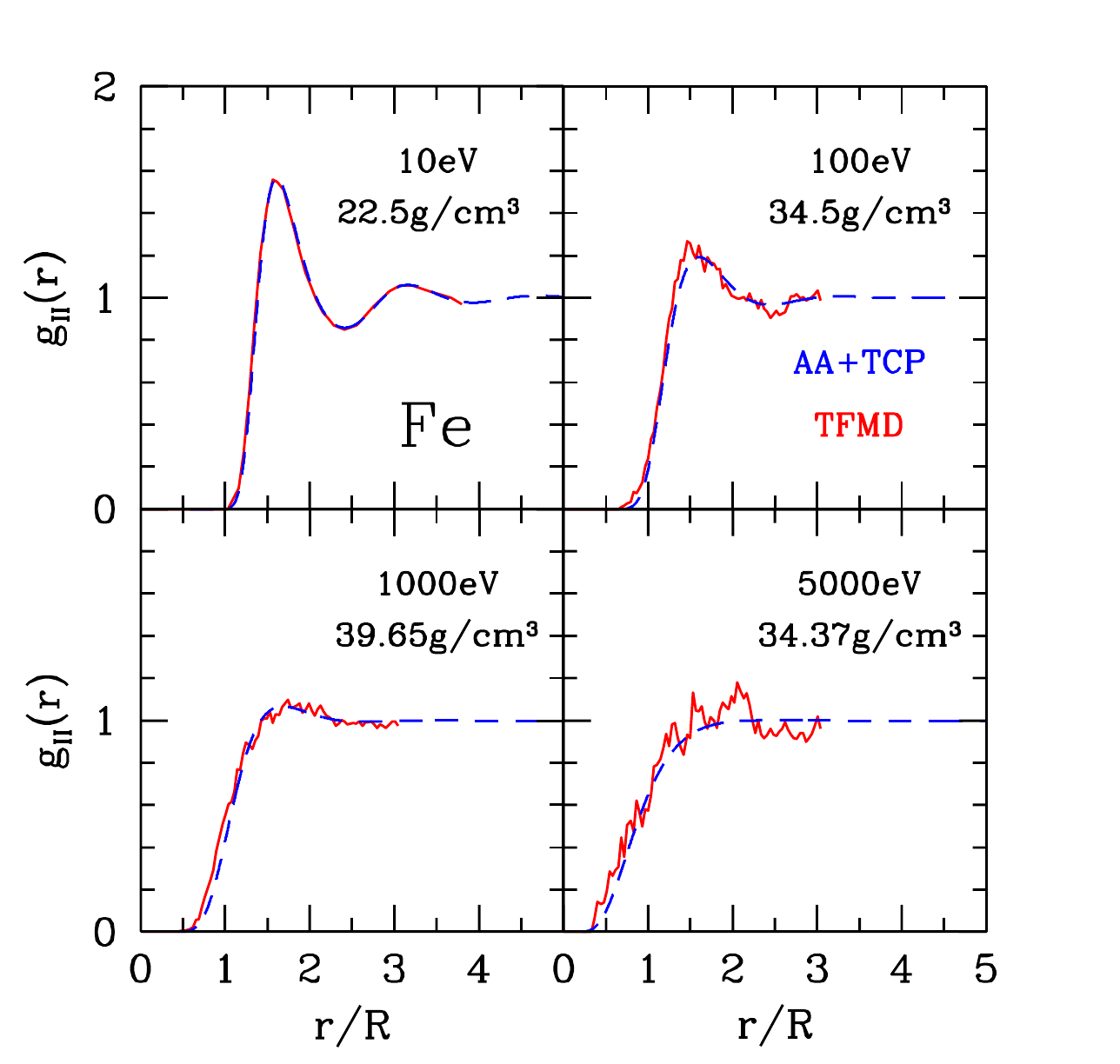}
  \caption{Pair distribution function of iron for four $(T,\rho)$ points along the principal Hugoniot.
           The red curves are Thomas-Fermi molecular dynamics (TFMD) simulations. The 10$\,$eV calculation 
           includes electron exchange \cite{kress_p} while the other three (100 -- 5000$\,$eV) 
           do not \cite{lambert3}.
           The AA+TCP calculations (blue) were done with the Thomas-Fermi model of the electrons 
           and all include exchange. $R$ is the ion sphere radius (Eq. \ref{Ris}).}
  \label{fig:Fe_g}
\end{figure}

For metals around solid density, QMD simulations become computationally impractical when the temperature is comparable to or 
larger than the Fermi temperature, which
typically puts an upper bound of $\sim 10-20\,$eV on the applicability of the method. However, the method can be extended to
higher $T$ by using the semi-classical Thomas-Fermi model of the electrons rather than a quantum model.
Such Thomas-Fermi Molecular Dynamics (TFMD) calculations have been performed along the principal Hugoniot of iron 
up to 5000$\,$eV \cite{lambert3} over compression ratios $\rho/\rho_0$ of 2.9 to 5.0. For this comparison (Fig. \ref{fig:Fe_g}) 
we ran the AA+TCP model with the Thomas-Fermi 
model of the electrons.  At $T=10\,$eV and 22.5 g/cm$^3$, both calculations include electron exchange and agree perfectly. The 
agreement remains excellent at the three higher Hugoniot points. For the latter, the TFMD calculations do not include electron exchange;
a contribution that should diminish rapidly as $T$ increases. The small shift in the highest temperature point (5000$\,$eV) for $r/R \wig< 1$
is somewhat puzzling given the lower ion-ion coupling and the excellent agreement at 1000$\,$eV.  On the other hand, a calculation with the
screened one-component plasma model (SOCP \cite{chabrier}), which uses linear response theory to describe the electron screening,
is in perfect agreement with our $g_{II}(r)$ at $T=5000\,$eV (not shown, as the curves are indistinguishable on the scale of the figure).
This suggests that the departure with the TFMD simulation at this very high temperature may be revealing some statistical inaccuracy in the
latter, as suggested by the growing level of noise in the simulations as $T$ increases.
The same calculations with the {\it quantum} AA+TCP give $g_{II}(r)$ that are identical to the Thomas-Fermi results \cite{starrett2}, with only
small differences appearing at 10$\,$eV. Thus the quantum version of the model recovers the Thomas-Fermi limit at high $T$ and
high $\rho$.

\begin{figure}
  \centering
  \includegraphics[scale=0.95, trim= 15mm 57mm 80mm 144mm, clip]{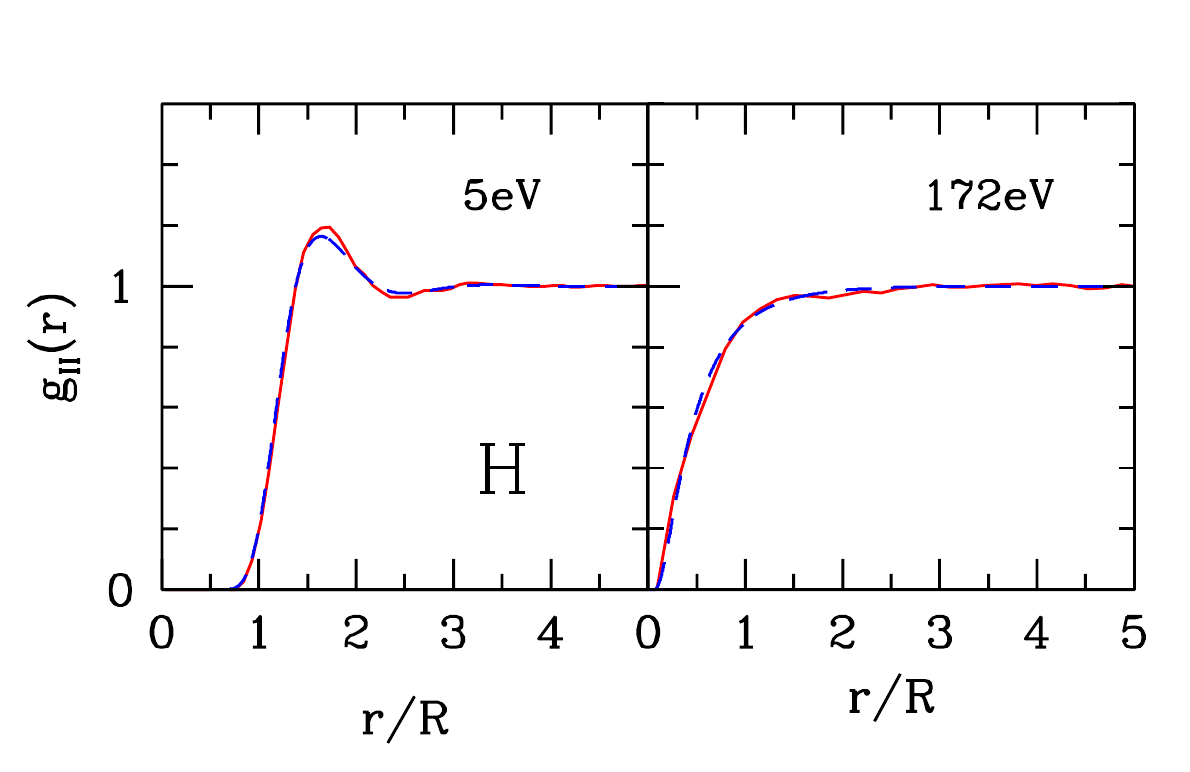}
  \caption{Pair distribution function of hydrogen at $80\,$g/cm$^3$ and $T=5$ and 172$\,$eV.
           The red curves are quantum molecular dynamics (QMD) simulations \cite{recoules3}.
           The results of the AA+TCP model are shown in blue. $R=0.323\,$a.u. is the ion sphere radius (Eq. \ref{Ris}).}
  \label{fig:H_g}
\end{figure}

Finally, the $g_{II}(r)$  of very dense hydrogen (80$\,$g/cm$^3$), or about 800 times the solid density) is found to agree perfectly 
(within the scatter) with QMD simulations \cite{recoules3} at $T=172\,$eV and to match very well the QMD result at 5$\,$eV (Fig. \ref{fig:H_g}). Again, a calculation
with the linear response SOCP model gives an identical $g_{II}(r)$ as with the AA+TCP model (not shown).
This demonstrates that in the limit of weak electron-ion coupling (but moderate to strong coupling and electron 
degeneracy), the quantum AA+TCP model recovers the proper limit of linear screening. 

While the model must be tested more extensively, several important points can already be made: 
\begin{enumerate}
  \item{The numerical solution of the model's equations has been successfully implemented and convergence achieved for a wide 
        range of warm and hot dense matter conditions ($Z$=1--26, $T=1-5000\,$eV, $\rho/\rho_0=1-800$).}
\item{Excellent agreement is found in comparisons of the pair distribution function with QMD/TFMD simulations.}
\item{The AA+TCP model with quantum electrons recovers the results of the AA+TCP with the Thomas-Fermi model of the electrons
      in the high-$T$, high-$\rho$ regime.}
\item{In the limit of weak ion-electron coupling, the quantum AA+TCP model recovers the SOCP model.}
\item{A Thomas-Fermi AA model with ion correlations can be successfully defined and implemented.}
\end{enumerate}

\section{Summary and outlook}
\label{conclusion}

We have presented a model of warm dense matter that extends the popular average atom (AA) models by introducing the correlations
in the surrounding plasma, which replaces a ``one ion'' description of WDM by one for the whole plasma. The
plasma is a mixture of classical ions and quantum electrons described with the 
integral equations theory of interacting fluids, which
we call a two-component plasma model (TCP). The AA
and TCP models are coupled self-consistently, each providing quantities necessary to close the other. The electronic structure surrounding
a nucleus takes into account Coulomb interactions with the surrounding plasma, whose ions and the response of the electron fluid 
(screening) are in turn described by the solution for the central ion. The resulting AA+TCP model has no
free parameters and only requires the nuclear charge $Z$, the temperature $T$, and the ion density $n_I^0$ as inputs. The
electrons can be treated semi-classically (Thomas-Fermi) or quantum mechanically (Schr\"odinger or Dirac equation). The form adopted for the coupling 
between the two models is original and leads to the recognition that the ion charge in the AA model is different from that
in the TCP model. The latter is the physically relevant charge related to observables. 
Earlier AA models with plasma correlations did not recognize this point and were overly constrained.

The introduction of plasma correlations increases considerably the mathematical and numerical complexity of the model,
which is mitigated by imposing spherical symmetry on the problem. The model was developed with an emphasis on formal
development, internal
consistency, and well-defined approximations, which we think, has been achieved to the extent that is possible when
combining an AA model with the integral equations theory of fluids. On the other hand, such a coupled model has intrinsic ambiguities
that require some ad hoc choices, such as in the criterion for the ideal chemical potential of the electrons and the 
definition of which electrons are to be counted as part of an ion. These are the less satisfying features of the model.
In principle, some of these choices must be better than others, something that can be established by comparison with
other theories that are free of such ambiguities (e.g. computer simulations, expansions around a non-interacting plasma state) 
or with experiments.

So far, the model has been validated by comparing the ion-ion pair distribution function of a wide range of warm and hot dense matter
systems with those calculated with {\it ab initio} simulations.  The agreement is uniformly excellent.  For cases
where the linear response describes the electron fluid accurately, comparisons with the 
screened one-component plasma model show perfect agreement. In those cases, deviations with the TFMD {\it ab initio} simulations shown suggest
that the latter may not be very accurate. The AA+TCP with a Thomas-Fermi model of the electrons is a viable model of
hot dense matter.  Finally, the AA+TCP model with quantum mechanical electrons recovers the 
AA+TCP model with the Thomas-Fermi model of the electrons at high densities and temperatures.  The goal of developing
a computationally efficient model has been achieved: Each of the results shown in Figs. \ref{fig:Al_g}--\ref{fig:H_g} 
runs in about one hour on a single processor workstation.

The realism of the AA+TCP model can be improved in several ways without modifying its structure (Fig. \ref{fig:aa_tcp_diag}) or
changing the key assumptions.  Foremost is introducing an ion-ion bridge function which will extend the accuracy of the model
to low temperatures where the coupling is very strong (e.g. liquid metals). Other possible refinements include the core overlap interaction 
in the direct ion-ion potential, relativity (Dirac equation), and a more sophisticated exchange and correlation 
potential, such as a generalized gradient approximation functional. 

Computer simulation methods combine theory and sophisticated algorithms to calculate the microscopic properties of dense plasmas.
The AA+TCP is a different approach that provides nearly all of the same microscopic properties.  Both methods can be thought of as
``central engines'' around which the calculation of many macroscopic properties of WDM can be built. The AA+TCP model can thus form the basis
for the calculation of nearly all the quantities typically obtained with computer simulations such as thermodynamics, 
conductivity, opacities, diffusion 
coefficients, and viscosity. It can be applied to the analysis of X-Ray Thomson scattering (XRTS) experiments as well as 
X-ray absorption near-edge spectroscopy (XANES) experiments. Furthermore, the model can be readily expanded to treat mixtures
of ions without any additional approximation or assumption.  Here the AA+TCP model offers a distinct advantage over simulations
as it can model highly asymmetric and very dilute mixtures.

This new average atom model with plasma correlations has so far shown a very satisfactory degree of physical realism.
It is a significant step beyond the more common ``atom in a cell'' models of warm dense matter. 
In view of its relatively modest computational cost and its many potential applications, it is a promising
approach to produce extensive tabulations of warm dense matter properties.

\begin{acknowledgement}
 We gratefully acknowledge V. Recoules, F. Lambert, J. D. Kress
 and L. Collins for providing pair distribution functions from their {\it ab initio} simulations.
 This work was performed under the auspices of the United States Department of Energy under contract DE-AC52-06NA25396.
\end{acknowledgement}
 
%
%
\bibliographystyle{/home/dsaumon/texmf/tex/latex/latex_packages/springer_book/styles/spphys}
\bibliography{bibfile}

\end{document}